\documentclass[usenatbib]{mn2e}

\pdfoutput=1

\usepackage{fixltx2e}
\usepackage{mathptmx}
\usepackage[pdftex]{graphicx}

\newcommand{\Mpc}{\rm\thinspace Mpc}
\newcommand{\kpc}{\rm\thinspace kpc}

\newcommand{\km}{\rm\thinspace km}

\newcommand{\cm}{\rm\thinspace cm}

\newcommand{\cmcu}{\hbox{$\cm^3\,$}}

\newcommand{\yr}{\rm\thinspace yr}

\newcommand{\s}{\rm\thinspace s}

\newcommand{\keVpcmcu}{\hbox{$\keV\cm^{-3}\,$}}

\newcommand{\Msun}{\hbox{$\rm\thinspace M_{\odot}$}}

\newcommand{\Msunpyr}{\hbox{$\Msun\yr^{-1}\,$}}

\newcommand{\keV}{\rm\thinspace keV}

\newcommand{\erg}{\rm\thinspace erg}

\newcommand{\ergps}{\hbox{$\erg\s^{-1}\,$}}

\newcommand{\kmps}{\hbox{$\km\s^{-1}\,$}}

\newcommand{\kmpspMpc}{\hbox{$\kmps\Mpc^{-1}$}}

\newcommand{\Zsun}{\hbox{$\thinspace \mathrm{Z}_{\odot}$}}

\newcommand{\psqcm}{\hbox{$\cm^{-2}\,$}}

\newcommand{\asec}{\rm\thinspace arcsec}

\begin{document}

\title[Multiple outbursts in 2A~0335+096]
{Feedback through multiple outbursts in the cluster 2A~0335+096}

\author[J.~S. Sanders, A.~C. Fabian \& G.~B. Taylor]
{J.~S. Sanders$^1$\thanks{E-mail: jss@ast.cam.ac.uk},
  A.~C. Fabian$^1$ and G.~B. Taylor$^{2,3}$\\
  $^1$ Institute of Astronomy, Madingley Road, Cambridge. CB3 0HA\\
  $^2$ Department of Physics and Astronomy, University of New
Mexico, Albuquerque NM, 87131, USA\\
  $^3$ Adjunct Astronomer at the National Radio Astronomy Observatory\\
}
\maketitle
  
\begin{abstract}
  We examine the core of the X-ray bright galaxy cluster 2A~0335+096 using deep \emph{Chandra} X-ray imaging and spatially-resolved spectroscopy, and include new radio observations. The set of around eight X-ray bright blobs in the core of the cluster, appearing like eggs in a bird's nest, contains multiphase gas from $\sim 0.5$ to 2~keV. The morphology of the coolest X-ray emitting gas at 0.5~keV temperature is similar to the H$\alpha$ emitting nebula known in this cluster, which surrounds the central galaxy. \emph{XMM-Newton} grating spectra confirm the presence of material at these temperatures, showing excellent agreement with \emph{Chandra} emission measures. On scales of 80 to 250~kpc there is a low temperature, high metallicity, swirl of intracluster medium as seen in other clusters. In the core we find evidence for a further three X-ray cavities, in addition to the two previously discovered. Enhancements in 1.5~GHz radio emission are correlated with the X-ray cavities. The total $4PV$ enthalpy associated with the cavities is around $5 \times 10^{59} \erg$. This energy would be enough to heat the cooling region for $\sim 5\times10^{7} \yr$. We find a maximum pressure discontinuity of 26 per~cent ($2\sigma$) across the surface brightness edge to the south-west of the cluster core. This corresponds to an upper limit on the Mach number of the cool core with respect to its surroundings of 0.55.
\end{abstract}


\section{Introduction}
We present deep \emph{Chandra X-ray observatory} data of the 2A~0335+096 cluster of galaxies also known as Zw~0335.1+0956. The cluster is 14th brightest in the sky in the 0.1--2.4~keV band from the sample of \cite{Reiprich02}. The X-ray source was first detected using \emph{Ariel 5} \citep{Cooke78} and is associated with the compact Zwicky poor cluster \citep{Schwartz80}. The cluster was inferred to have a short cooling time from its X-ray properties, a result confirmed by later \emph{ROSAT} observations \citep{Sarazin92}. More recently \emph{Chandra} has observed a great deal of structure in its core \citep{Mazzotta03}. They found around eight cool dense blobs, each around 4~kpc in radius, and two cavities. The deprojected temperature of these blobs was claimed to show that they may not be in thermal pressure equilibrium with their surroundings. The cluster contains a cold front 40~kpc to the south of the nucleus. The discontinuity in the thermal pressure across this front implies the core has Mach number of $\sim 0.75$ relative to its surroundings. \cite{SandersShock06} instead put forward the idea that the cold front may be an isothermal shock as found in the Perseus cluster.

\cite{Werner06} examined the blobs in \emph{XMM-Newton} data, finding a lower temperature for them than \cite{Mazzotta03}. They instead claimed that the blobs are in pressure equilibrium with their environment. They also examined the relative amounts of different metals, computing a relative number contribution of Type Ia supernovae of 25 per cent.

The galaxy cluster contains a complex radio source \citep{Sarazin2A033595}. Twin jets emerge from both sides of a compact radio nucleus embedded in the dominant cluster galaxy, and extend $\sim 12$~arcsec in length.   The nucleus of a nearby companion galaxy to the northwest is also detected in the radio as a point source. Both galaxies are embedded in an extended, steep spectrum mini-halo present in the central $\sim 100$~kpc of the cluster.

The galaxy cluster contains a complex radio source \citep{Sarazin2A033595}. There are two lobes or jets are either side of the nucleus, around 12 arcsec in size. There is also a bright extended minihalo component with a steep spectrum. The central galaxy and its nearby companion to the north west are also detected as radio point sources.

The centre of the cluster also hosts a complex filamentary emission-line nebula \citep{Romanishin88,Donahue07}, around 30 arcsec in size. The luminosity of the nebula in H$\alpha$+[N~\textsc{ii}] is similar to the famous nebula in the Perseus cluster \citep{Donahue07,Minkowski57}. The velocities of the central galaxy and its companion suggest they may be merging \citep{Gelderman96,Donahue07,Hatch07}.

2A~0335+096 lies at a redshift of 0.0349. Therefore 1 arcsec on the sky corresponds to a distance of 0.7~kpc on the object, assuming a Hubble constant of $70 \kmpspMpc$, which gives an angular diameter distance of 143~Mpc. In each of the images shown in this paper, north is to the top and east is to the left.

\section{Data analysis}

\begin{table}
  \caption{\emph{Chandra} observations analysed in this paper. The filtered column shows the observation exposure after filtering for flares.}
  \begin{tabular}{llll}
    \hline
    Date       & Observation ID & Exposure (ks) & Filtered (ks) \\
    \hline
    2000-09-06 & 919            & 19.7          & 10.4 \\
    2007-12-16 & 7939           & 49.5          & 49.5 \\
    2007-12-20 & 9792           & 33.7          & 33.7 \\
    Total      &                &               & 93.7 \\
    \hline
  \end{tabular}
  \label{tab:obs}
\end{table}

2A~0335+096 was observed three times by \emph{Chandra} (Table \ref{tab:obs}). Each observation was made using the ACIS-S3 detector aimed at the centre of the cluster. To clean the observations for flares, we examined the 2.5 to 7~keV band using data from the ACIS-S1 detector.  Our two latest observations (7939 and 9792) were free from flares, so we did not exclude any time periods from these. The 919 observation contains a mild flare, as found by \cite{Mazzotta03}, which we removed from that dataset.  The total exposure time of the observations, after cleaning for flares, is 93.7~ks.

We applied the latest processing to each of the datasets using \textsc{acis\_process\_events}. The GAINFILE used was acisD2000-01-29gain\_ctiN0006.fits, the CTIFILE acisD2002-08-01ctiN0006.fits and the TGAINFIL was acisD2007-11-01t\_gainN0006.fits. The 919 and 9792 event files were reprojected to match the 7939 observation.

\begin{figure}
  \centering
  \includegraphics[width=0.8\columnwidth]{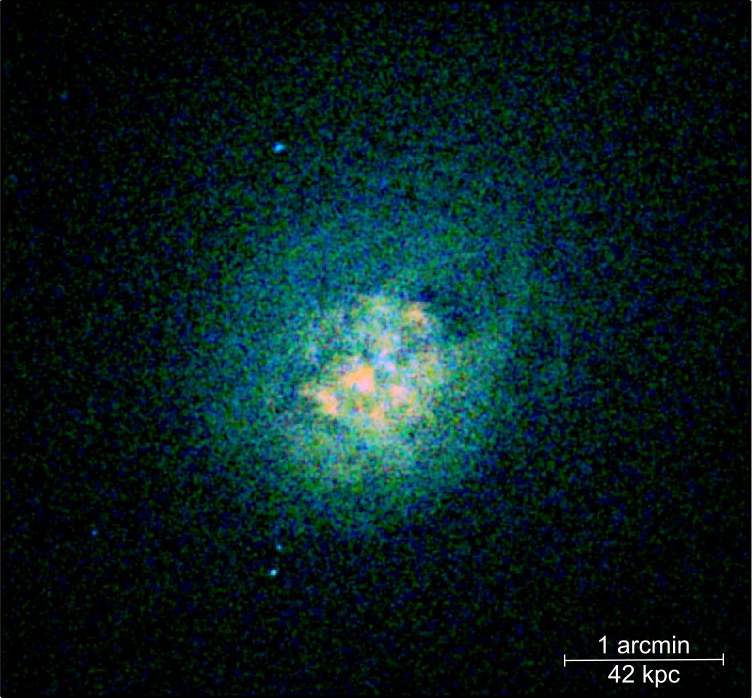}\\
  \includegraphics[width=0.8\columnwidth]{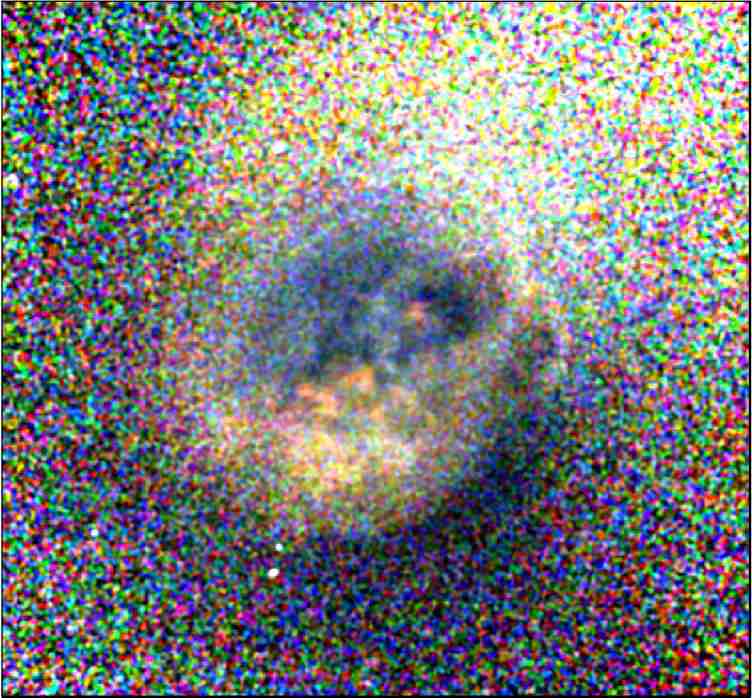}
  \caption{Colour images. The band 0.3--1.2, 1.2--2.5 and 2.5--7.0~keV are shown in red, green and blue, respectively. The top panel shows an unsharp-masked enhanced image of the cluster (see text). The bottom panel shows an image showing the fractional deviation in each band relative to the average brightness at each radius.}
  \label{fig:colimages}
\end{figure}

\section{X-ray imaging analysis}
We show in Fig.~\ref{fig:colimages} red-green-blue (RGB) images of the cluster core.  These were created in the 0.3--1.2, 1.2--2.5 and 2.5--7.0~keV energy bands. The count images were exposure-corrected by the total 1.5~keV monochromatic exposure map for the observations. The upper panel shows the RGB image generated by unsharp-masking the individual energy bands. We Gaussian-smoothed each image by 1 and 10~arcsec, subtracting 70 per cent of the 10-arcsec-smoothed image from the 1-arcsec-smoothed image. The unsharp masking enhances the smaller-scale features. The bottom panel was created with a different technique, where in each band we calculate for each pixel the fractional difference between the brightness of that pixel and the average at each radius (measuring the centre from the position of the radio source, which lies close to the centre of the images).

The centre of the cluster has a complex structure, consisting of a number of X-ray bright knots emitting soft X-ray emission \citep{Mazzotta03} and surface brightness depressions.

\begin{figure}
  \centering
  \includegraphics[width=0.8\columnwidth]{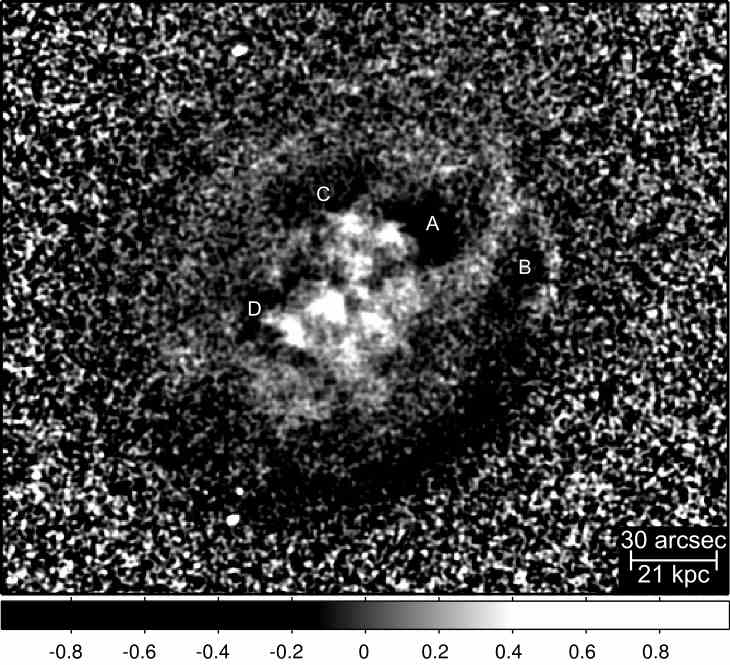}
  \caption{Unsharp-masked image, showing the fractional difference between an image smoothed by a Gaussian of 1 pixel width and one smoothed by a Gaussian of 30 pixels. The resultant image was finally smoothed by a Gaussian of 2 pixels. The X-ray cavities are labelled A--D.}
  \label{fig:unsharp}
\end{figure}

As in other clusters, the surface brightness depressions are likely to be X-ray cavities generated by the displacement of X-ray emitting gas by nonthermal plasma (see \citealt{McNamaraNulsen07} for a review). These depressions are easier to see in a total unsharp-masked image Fig.~\ref{fig:unsharp}, where we label the strongest decrements as A--D. Depressions A and D were seen in the earlier \emph{Chandra} observation \citep{Mazzotta03}. Depressions B and C are new to this deeper \emph{Chandra} dataset. It is possible that A and B are actually part of a larger bubble, most easily seen by examining the lower panel of Fig.~\ref{fig:colimages}, where it appears that the western rim of cavity B could join onto the eastern rim of cavity A.

The lower panel of Fig.~\ref{fig:colimages} also shows that there is excess emission to the north-west of the cluster core relative to other directions. We also see the sharp drop in surface brightness at 1 arcmin radius to the south-east of the core, identified as a cold front in \cite{Mazzotta03}.

\begin{figure}
  \centering
  \includegraphics[width=\columnwidth]{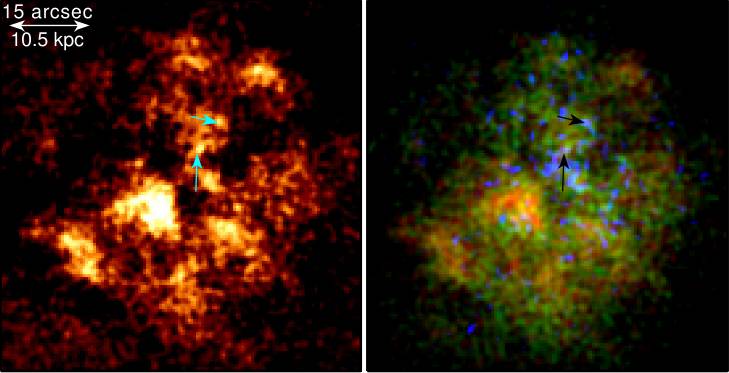}
  \caption{(Left panel) Unsharped-masked detail of the centre of the cluster, created by subtracting an image smoothed by a Gaussian of 30 pixels from image smoothed by a Gaussian of 1 pixel. The location of the nuclei of the galaxies are marked by arrows. (Right panel) RGB detail of the centre of the cluster using the same energy bands as Fig.~\ref{fig:colimages}.}
  \label{fig:centredetail}
\end{figure}

We examine the very central region in more detail in Fig.~\ref{fig:centredetail}. Presented in the left panel is a full-band unsharp masked image of the central $45\times45$~kpc. Marked by arrows are the locations of the central D galaxy and its companion. In the rightmost panel is a detailed RGB image of the core without unsharp masking. There is considerable colour variation over this region. Most of the X-ray bright features are spectrally soft, being redder in the image. The regions close to the galactic nuclei are harder than their surroundings. This may be due to absorbing material associated with the galaxies. There are also faint point sources close to the location of the nuclei, much fainter than the surrounding gas.

\section{Spatially-resolved X-ray spectroscopy}
To examine the properties of the intracluster medium in more detail we applied spatially-resolved spectroscopy techniques. We split the data into spatial bins using the Contour Binning algorithm of \cite{SandersBin06}. This procedure chooses regions with a certain signal-to-noise ratio, but follows directions of equal surface brightness, matching the morphology of the X-ray emission.

For each of the \emph{Chandra} observations we extracted spectra from each of these regions. We added the spectra together using the \textsc{mathpha}, part of the \textsc{ftools} package. We generated emission-weighted response and ancillary response files using the \textsc{ciao} tools \textsc{mkacisrmf} and \textsc{mkwarf}, respectively. We used \textsc{ciao} version 4.0 and \textsc{caldb} version 3.5.0. The responses were calculated weighting the calibration regions according to the number of counts between 0.5 and 7~keV. The response files from each of the observations were averaged together, weighting according to the number of counts in the respective observations.

The spectra were fit in \textsc{xspec} 12.5 \citep{ArnaudXspec} using the \textsc{apec} spectral model \citep{SmithApec01}. We fit the spectra between 0.5 and 7~keV. In the spectral fitting we froze the absorbing column density at the average best-fitting value obtained from fitting regions away from the cluster centre, $2.4 \times 10^{21} \psqcm$. We note that this value is considerably higher than the value of $1.75\times 10^{21} \psqcm$ obtained from H\textsc{i} surveys \citep{Kalberla05}. At the high Galactic column density of the cluster, it is likely there is absorbing structure on scales smaller than the $0.5\deg$ sampling of the H\textsc{i} survey. Large absorbing column densities of $2.5\times 10^{21} \psqcm$ have been observed previously by \emph{ROSAT} \citep{Irwin95} and more recently by \emph{XMM-Newton} \citep{Werner06}, justifying freezing the spectral fits at our chosen value.

We generated background spectra from the standard blank-sky backgrounds. We split up the appropriate ACIS-S3 background\footnote{acis7sD2000-01-29bkgrnd\_ctiN0001.fits} into three sections to make a background for each of the observations analysed. Each section had the same ratio of exposure time to the total background exposure time, as each foreground observation had to the total foreground. We normalised each background observation to match the 9 to 12~keV particle and instrumental count rate for the respective foreground observation. Finally, we reprojected the backgrounds to match the original \emph{Chandra} observations, then reprojected them to match the 7939 observation.

By matching the hard background rates, we may over or underestimate the soft X-ray background. To check that this was not a significant problem, we also generated ACIS-S1 backgrounds files using standard blank sky backgrounds. The ACIS-S1 CCD has the same back-illuminated detector technology as the ACIS-S3 but does not contain bright cluster emission. The normalised background spectra closely matched the ACIS-S1 spectra from our observations over the whole energy band, showing that this is not an issue.

\begin{figure}
  \centering
  \includegraphics[width=0.8\columnwidth]{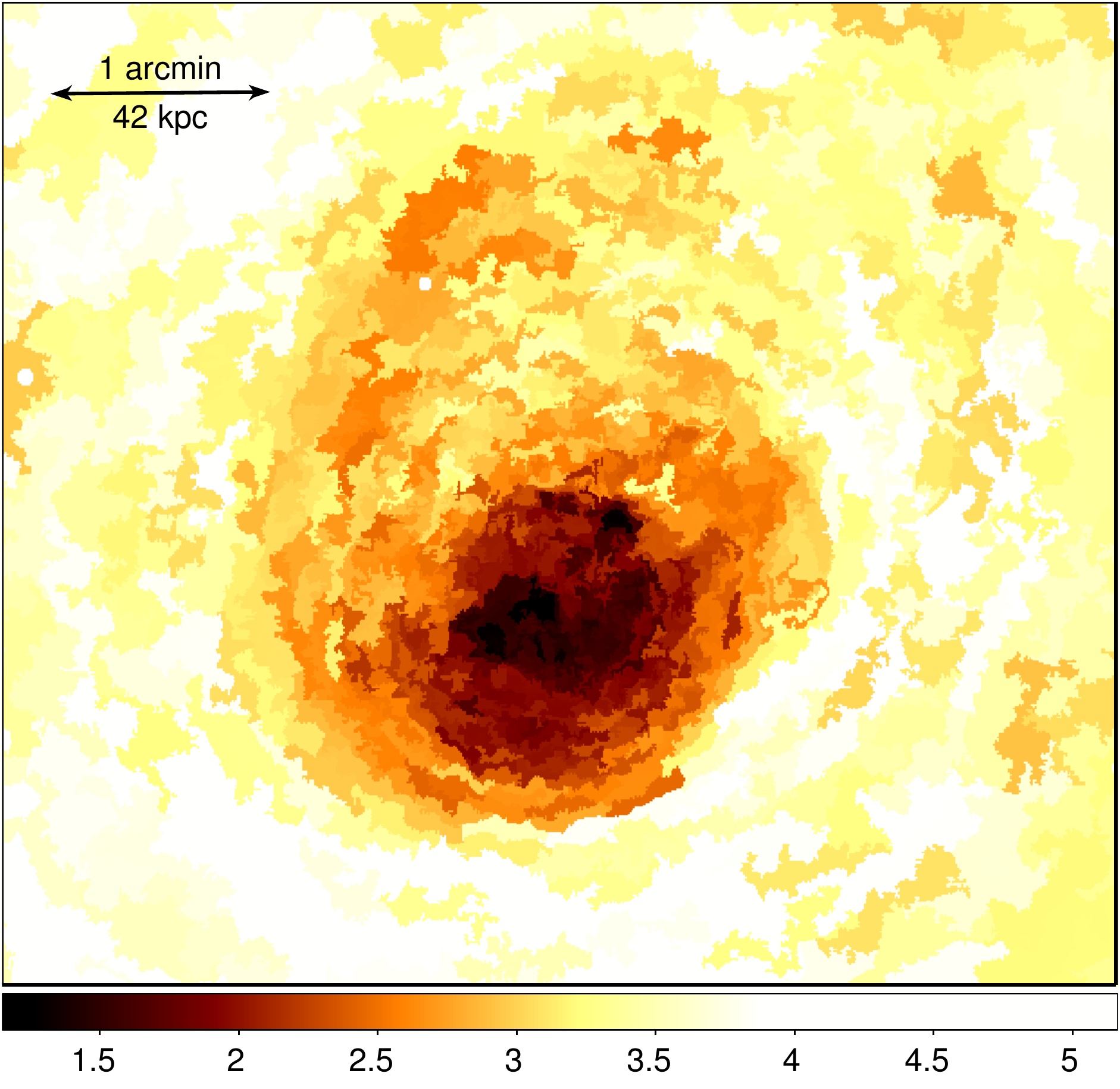}\\
  \includegraphics[width=0.8\columnwidth]{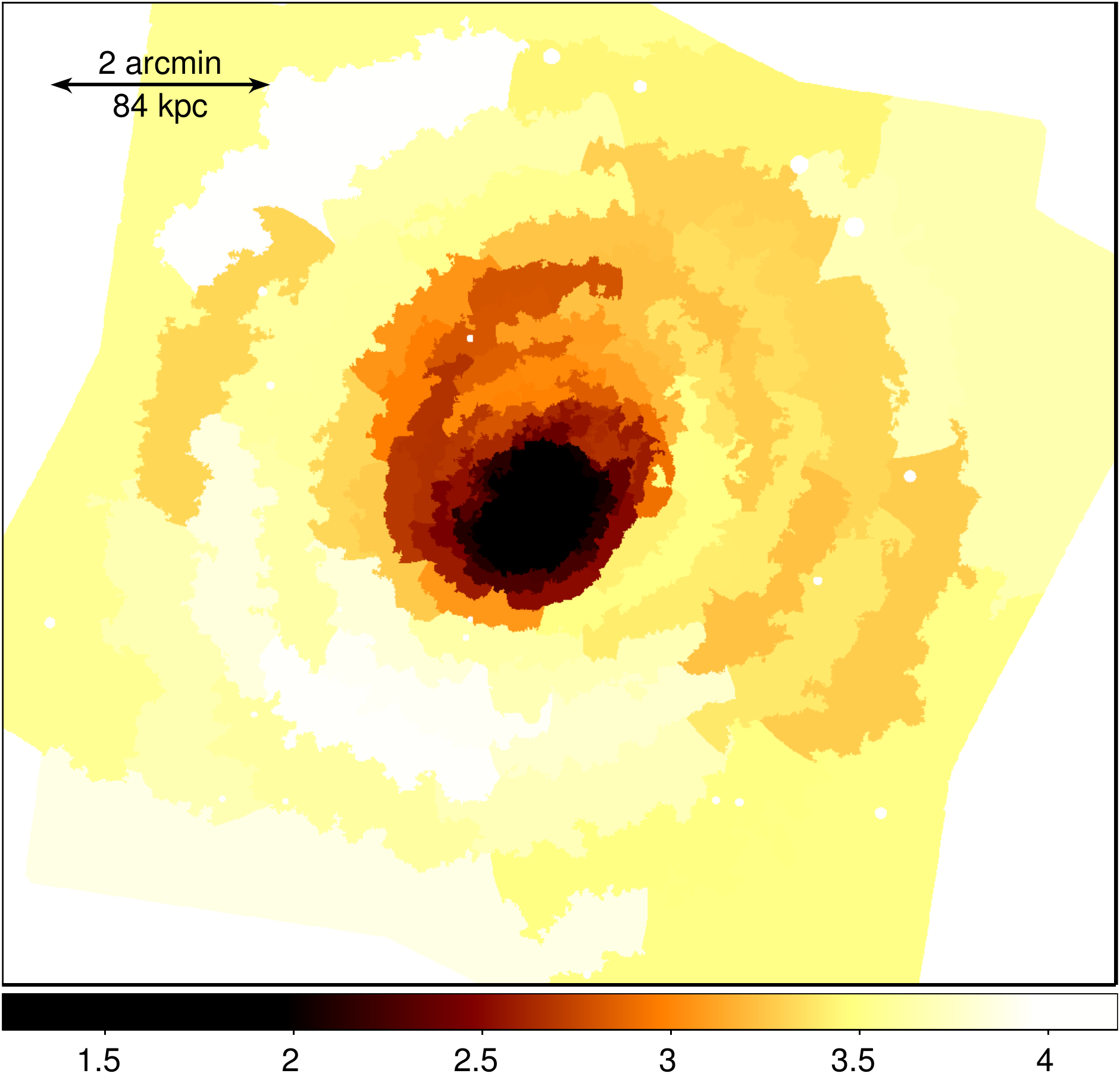}
  \caption{Temperature maps derived from spectral fitting of spectra extracted from regions. The top panel shows uses bins containing $\sim 1000$~counts, while the bottom panel uses bins containing 10\,000 counts. The scale is in keV.}
  \label{fig:tmaps}
\end{figure}

In Fig.~\ref{fig:tmaps} we show maps of the emission-weighted projected temperature. The detailed map of the central region in the top panel was created fitting spectra with a signal-to-noise ratio of $>32$ ($\sim 1000$ counts). In the spectral fits the metallicity, temperature and emission measures were allowed to be free. The C-statistic was minimised in the spectral fitting. The temperature measurements are accurate to 3 per cent in the coolest regions to around 7 per cent at the edge of the map.

The bottom panel of Fig.~\ref{fig:tmaps} shows the projected temperature on larger scales using regions with a signal-to-noise ratio of greater than 100 ($\sim 10^{4}$ counts). In these fits we minimised the $\chi^2$ statistical.  The statistical uncertainties of the temperature range from 0.8 per cent to 4 per cent.

The temperature maps show there are a number of cool blobs with projected temperatures of around 1.3~keV embedded in a core around 2~keV in temperature. The cool blobs correspond to the knots of soft emission seen in the colour images (Figs. \ref{fig:colimages} and \ref{fig:centredetail}). Outside of the cool core is a spiral-like plume of cooler X-ray emission gas, around 1~keV cooler than the surrounding projected temperature.

\begin{figure*}
  \includegraphics[width=0.3\textwidth]{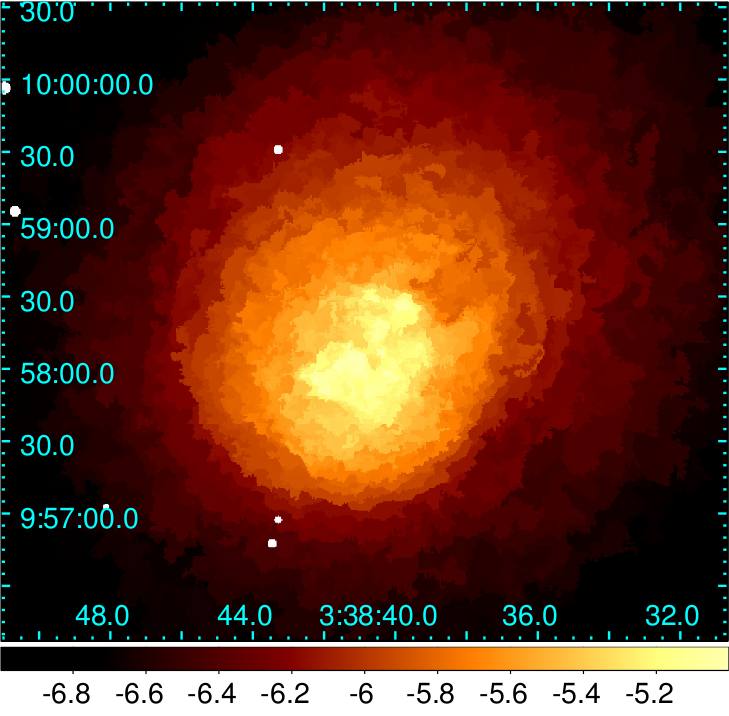}
  \includegraphics[width=0.3\textwidth]{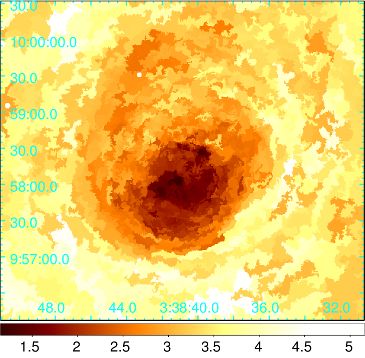}
  \includegraphics[width=0.3\textwidth]{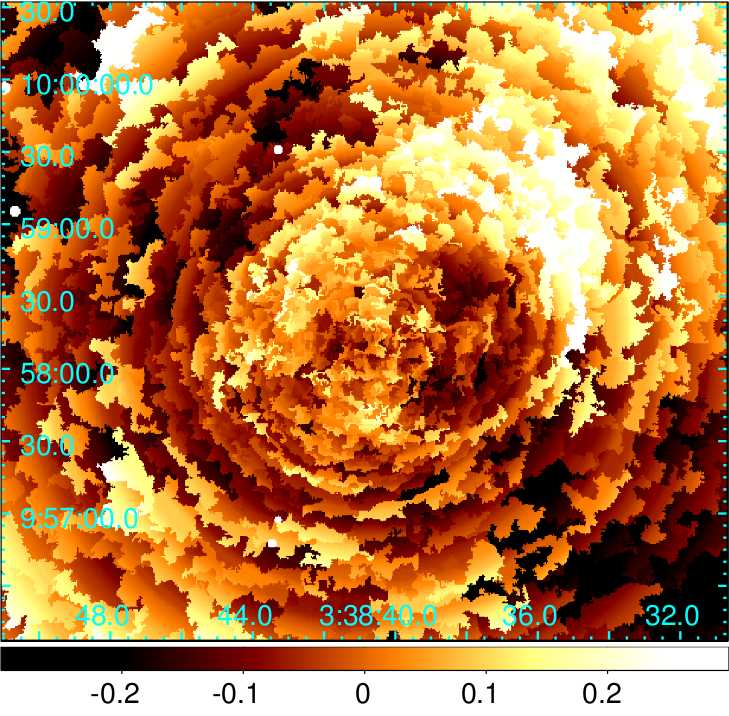}
  \caption{(Left panel) Emission measure per unit area (units are $\log_{10} \cm^{-5} \asec^{-2}$). (Centre panel) Temperature (keV). (Right panel) Projected pressure relative to average at each radius, constructed from square root of the emission measure per unit area times the temperature.}
  \label{fig:norm_temp_press}
\end{figure*}

Assuming a single phase, $P_\mathrm{proj} = kT \sqrt{\mathcal{E}}$, where $kT$ is the temperature and $\mathcal{E}$ is the emission measure per unit area, is a thermal pressure-like quantity, neglecting the line-of-sight depth. Fig.~\ref{fig:norm_temp_press} shows the best fitting emission measure per unit area and temperature, and the calculated projected pressure, relative to the average at each radius (helping remove geometrical effects). These quantities were derived using regions with a signal to noise ratio of at least 32. The pressure image has a surprising amount of structure within the core region and is not simply a reflection of the surface brightness or temperature.

Typical deviations of pressure relative to the average are around 10 per cent. Interesting features include an approximately circular region of lower thermal pressure (centred on 03:38:38.4, +09:59:20 and 45 arcsec in radius), with an enhancement of pressure around its edge (particularly to the south west). This circular region may be associated with cavity A. Fig.~\ref{fig:norm_temp_press} also shows that the cool spiral-shaped plume of gas has a lower thermal pressure than material at the same radius, though this is likely to be due to the shorter line-of-sight through the plume of the dominant material.

\begin{figure}
  \centering
  \includegraphics[width=0.6\columnwidth]{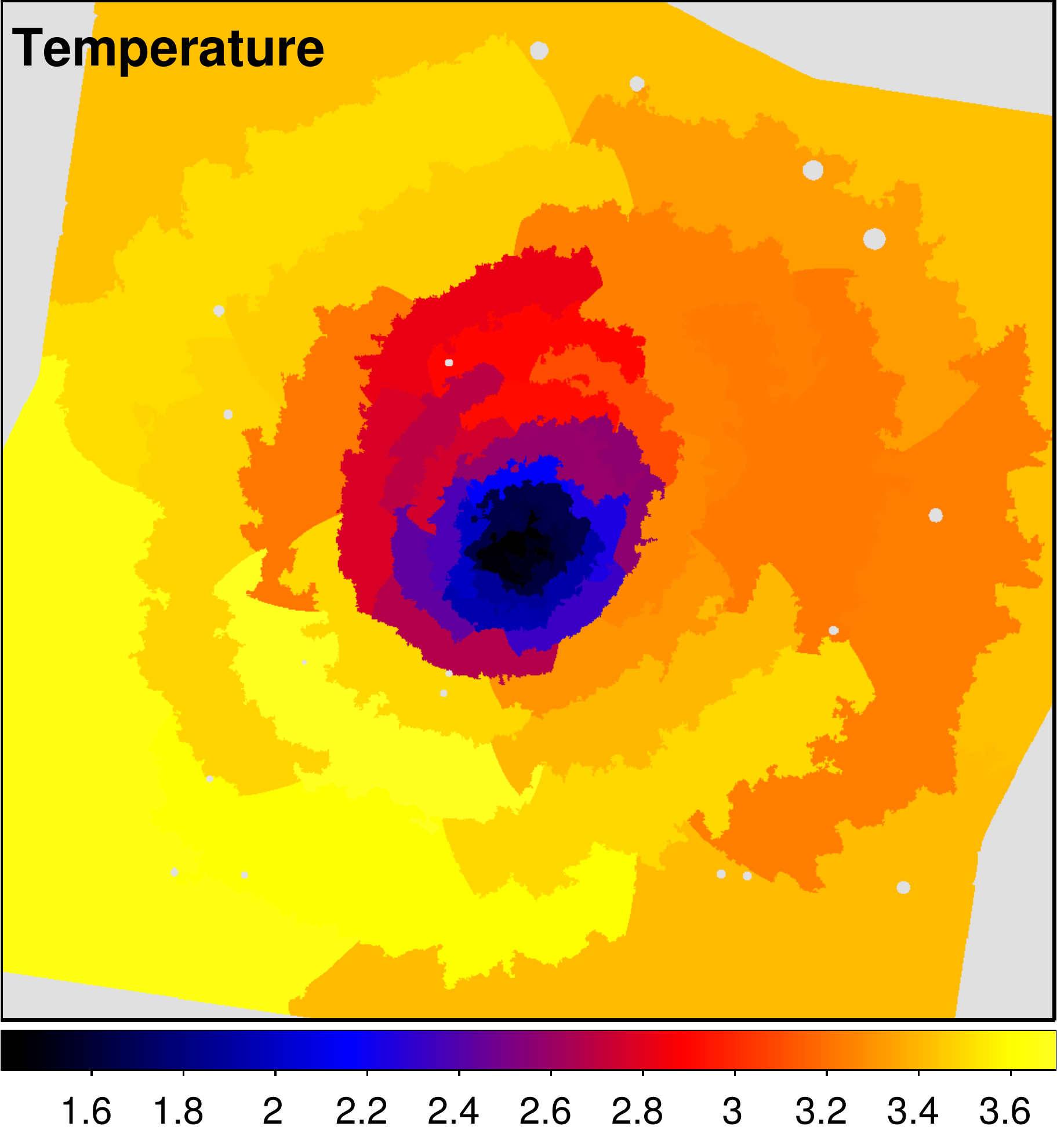}\\
  \includegraphics[width=0.6\columnwidth]{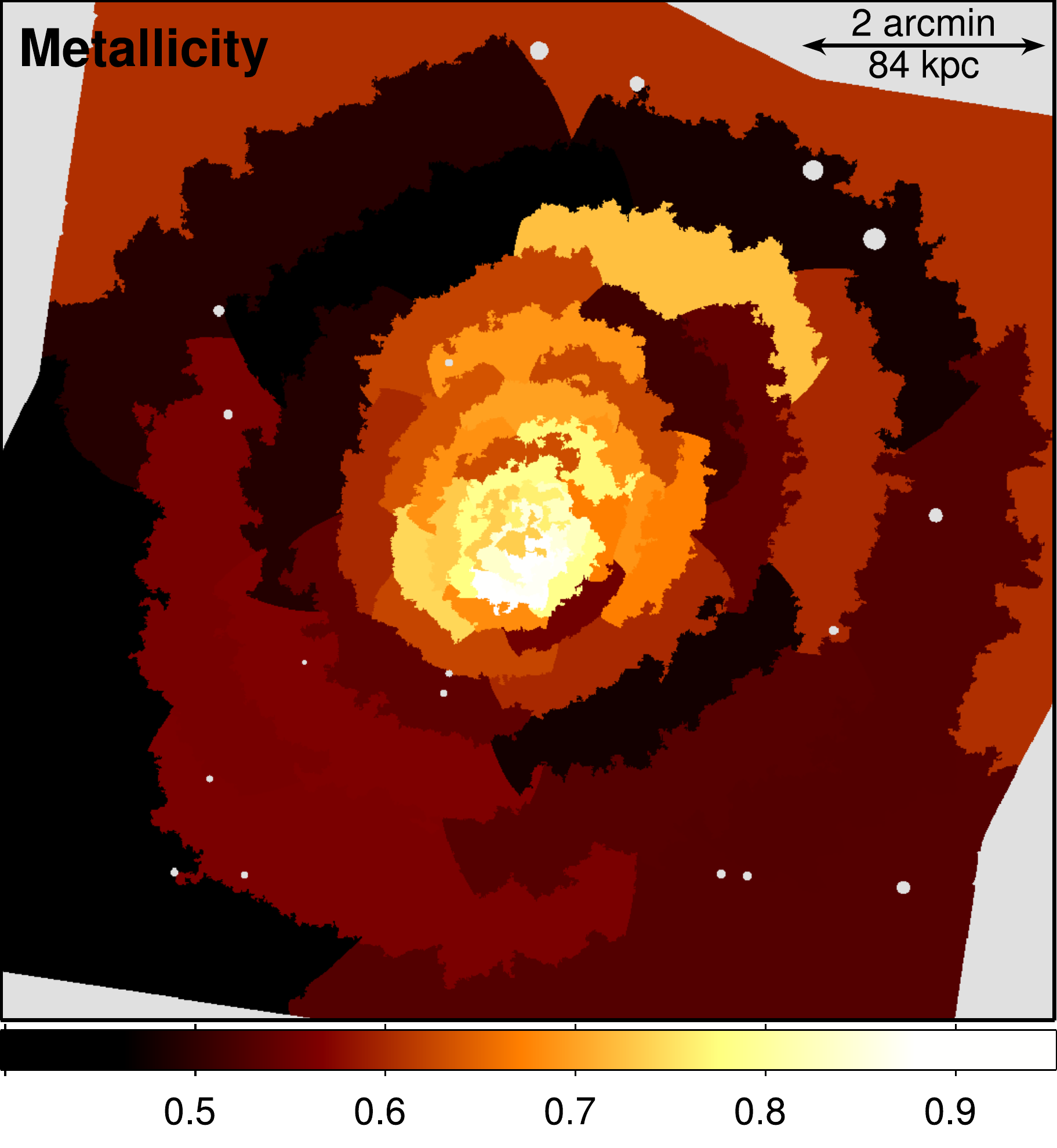}
  \caption{Comparison of metallicity and temperature from spectral fitting. The regions shown were chosen to have a signal to noise ratio greater than 150 (around 22\,500 counts). The temperatures are measured in keV with statistical uncertainties ranging from 0.01 keV in the centre to 0.16~keV in the outskirts using an isothermal spectral model. The metallicities are measured relative to Solar, measured using an multi-temperature spectral model, with uncertainties ranging from 0.04 to $0.09 \Zsun$.}
  \label{fig:zmap}
\end{figure}

We examine the metallicity structure in Fig.~\ref{fig:zmap}, also showing a temperature map with the same binning for reference. We used regions with a minimum signal to noise ratio of 150 (around 22\,500 counts). The temperature map just uses a single-temperature model fit, so it shows the mean projected emission-weighted temperature. The metallicity map, however, was created using a multi-temperature spectral model with components at 0.5, 1, 2, 4 and 8~keV, with fixed column density. A multi-temperature model is necessary because of the multiple temperature components necessary to fit the spectra in the centre. If a single temperature model is used the metallicity appears to decline in the core (\citealt{Mazzotta03} found a declining metallicity). In the outskirts, the metallicity is close to what is measured by a single temperature model. The metallicity from the multi-temperature model peaks at $0.95\pm 0.06 \Zsun$ at the centre.

We also examine the temperature, density and pressure of the intracluster medium across the cold front to the south west of the cluster core. We created 12 annuli centred on 03:38:41.29, +09:57:56.8 between angles of 156.4 to $261.2\deg$ east from north. Note that this is not the same range of angle as used by \cite{Mazzotta03} --- they used from 150 to $210\deg$ --- but we found the edge is not very sharp over this entire range. We use the same range of angle as in \cite{SandersShock06}.

\begin{figure}
  \centering
  \includegraphics[width=0.8\columnwidth]{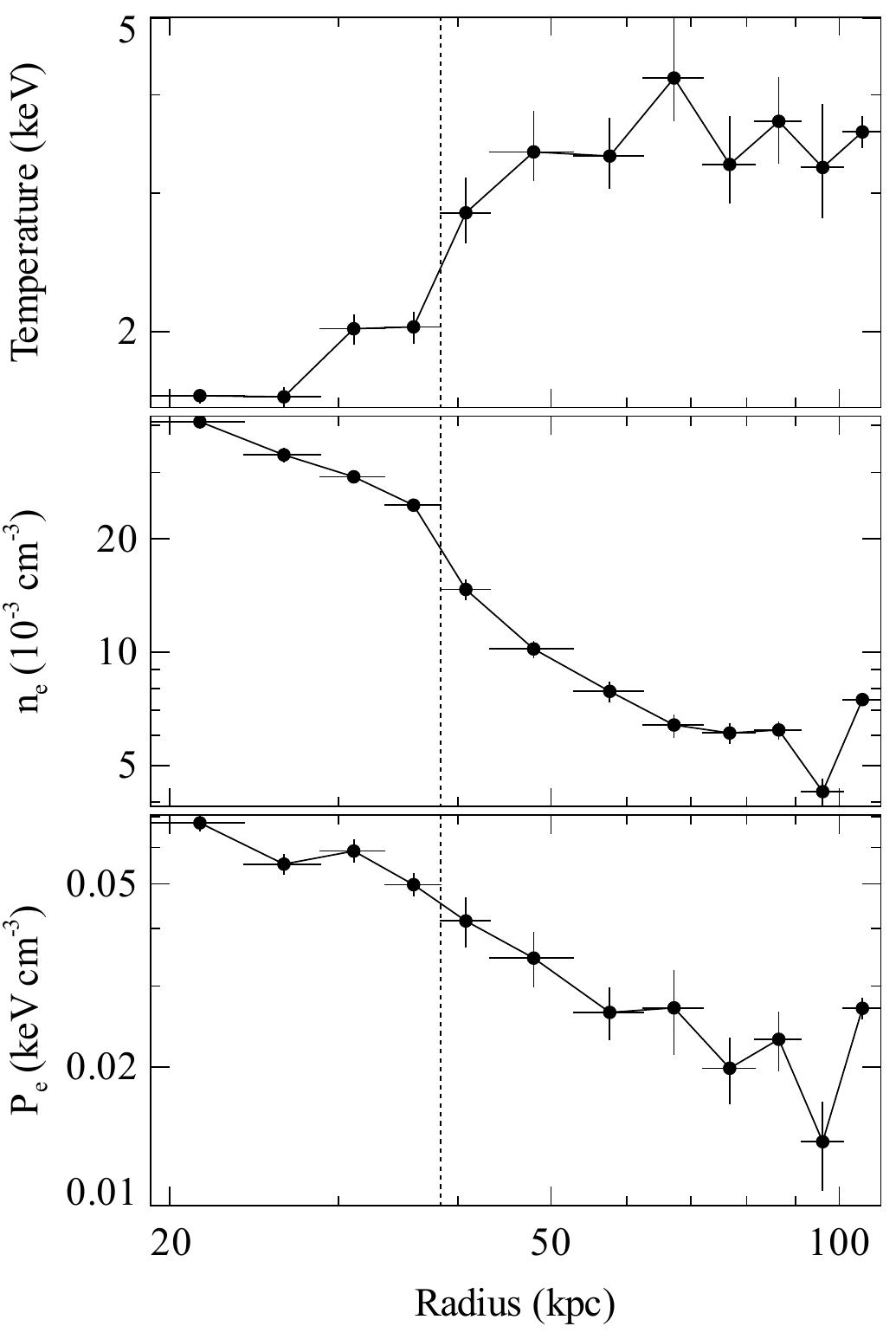}
  \caption{Deprojected temperature, electron density and electron pressure profiles across the cold front to the south west. The dotted line marks the radius of the cold front.}
  \label{fig:coldfront}
\end{figure}

We extracted projected spectra and background spectra from these regions and deprojected them with the deprojection method of \cite{SandersPer07} (\textsc{dsdeproj}; described more fully and tested in \citealt{Russell08}). We fitted the spectra with the \textsc{apec} model, allowing the metallicity, temperature and normalisation to be free, but fixing the column density at the Galactic value. The deprojected temperature profile is shown in Fig.~\ref{fig:coldfront}. Also shown are the electron densities calculated from the emission measure, and the electron pressure, computed assuming the uncertainties in temperature and density are independent. The dotted line marks the radius of the cold front. The results show that the pressure is continuous across the front.

To look for multiphase material in the cluster we fitted a model made up of several different  components at fixed temperatures and allowed the quantity of each component to vary in the fit. The spectral model contained components at 0.5, 1, 2 and 4 keV, fixing the metallicity at $0.8\Zsun$ and the column density at the Galactic value. The data were binned into regions containing a signal-to-noise ratio of 20 or greater ($\sim 200$ counts). The C statistic was minimised in the spectral fitting.

\begin{figure}
  \centering
  \includegraphics[width=\columnwidth]{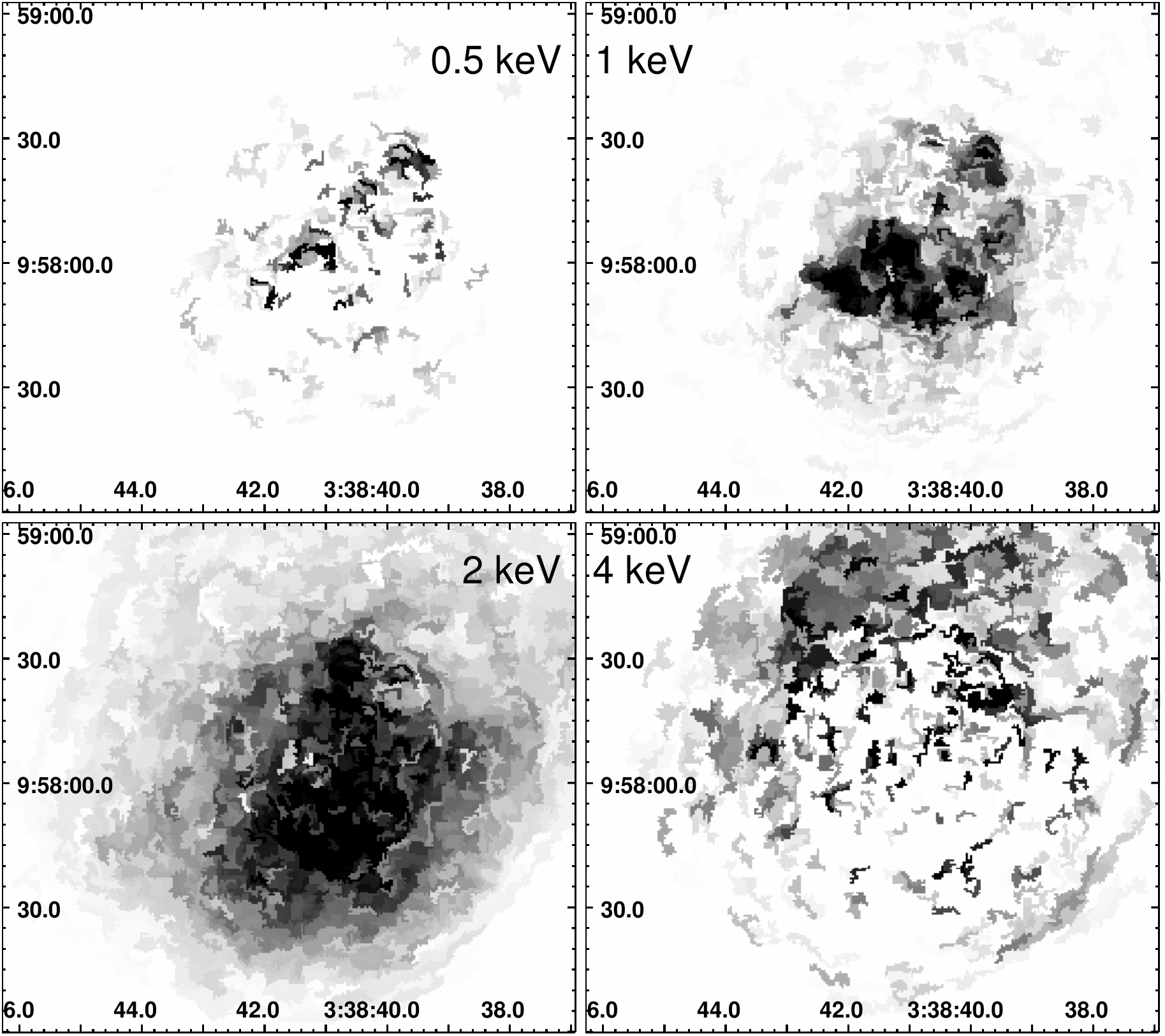}
  \caption{Emission measures per unit area from multi-temperature X-ray spectral fitting. Shown are the emission measures for the 0.5~keV component (top-left panel), 1~keV component (top-right panel), 2~keV component (bottom-left panel) and 4~keV component (bottom-right panel).}
  \label{fig:multitemp}
\end{figure}

\begin{figure}
  \centering
  \includegraphics[width=0.8\columnwidth]{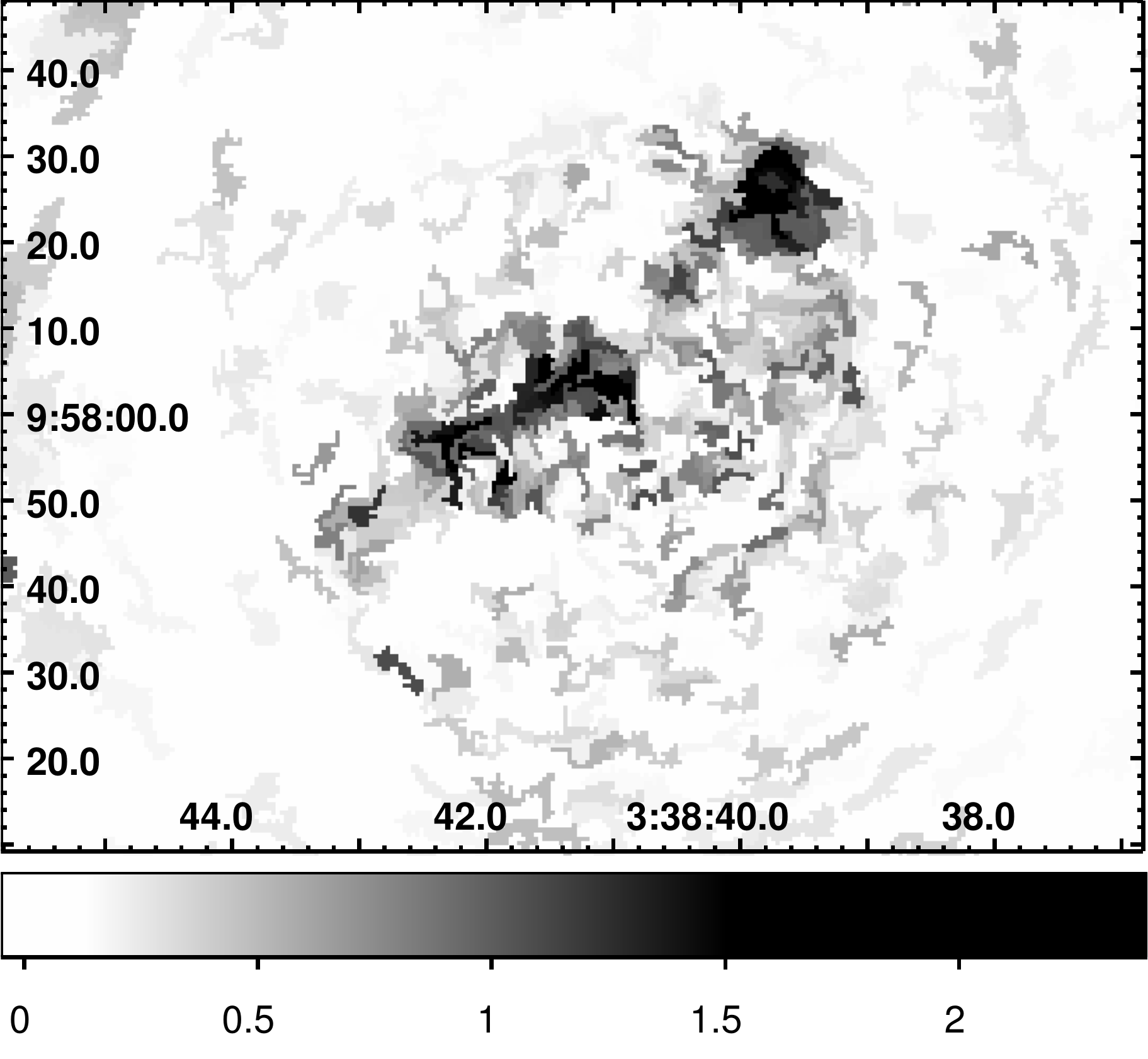}
  \caption{Decrease in reduced $\chi^2$ from single temperature model to multi-temperature model.}
  \label{fig:multideltachi2}
\end{figure}

Fig.~\ref{fig:multitemp} shows the emission measure of each component per unit area on the sky. The results show that the cool blobs in the centre of the cluster contain a range in temperatures, from $\sim 0.5$ to 2~keV. We can ask whether the multiple temperature components are necessary for the spectral fitting. Fig.~\ref{fig:multideltachi2} shows the reduction in reduced $\chi^2$ in the spectral fits from a single temperature model to a four component multi-temperature model. Note that this shows the results from $\chi^2$ fitting, not C statistic fitting. The image shows that those regions with strong 0.5 keV components have much poorer single temperature fits than multi-temperature fits, so the multiple components are necessary.

\begin{figure}
    \centering
    \includegraphics[width=0.9\columnwidth]{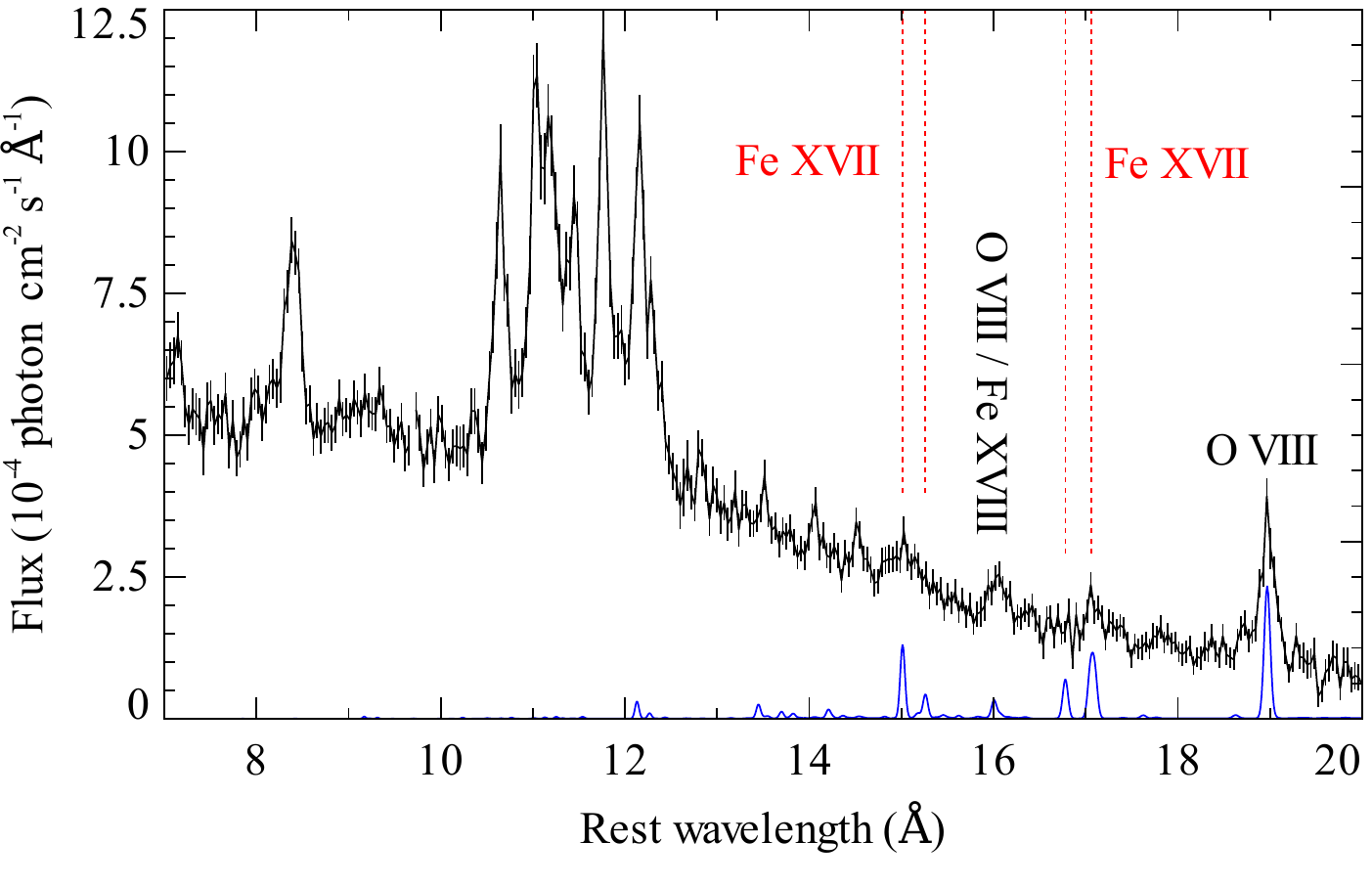}
    \caption{Fluxed \emph{XMM-Newton} RGS spectrum of the core of the cluster, combining the first and second order spectra from both RGS instruments. Shown are the positions of the Fe~\textsc{xvii} emission lines. A 0.4~keV model spectrum is shown for reference as the solid line beneath the data.}
    \label{fig:rgsspec}
\end{figure}

The amount of cool gas can be better quantified by using high spectral resolution \emph{XMM-Newton} Reflection Grating Spectrometer (RGS) data. The RGS spectra from this cluster shows Fe~\textsc{xvii} emission lines (\citealt{Werner06}; Fig.~\ref{fig:rgsspec}), indicating material around $\sim 0.5$~keV.

To examine the temperature distribution quantitatively we examined \emph{XMM-Newton} dataset 0147800201. We processed the data with \textsc{rgsproc} in \textsc{sas} 8.0.0. We filtered flares by excluding bad time periods where the count rate exceeded 0.2 counts per second on CCD 9 at an absolute cross dispersion angles greater than $1.5 \times 10^{-4}$ (using flag values of 8 and 16), giving a total exposure of 113~ks for each of the instruments. We extracted the spectra including 90 per cent of the point spread function (PSF) and 95 per cent of the pulse height distribution, to concentrate on the core of the cluster. The observation has a position angle of $73.124^\circ$, which is the angle of the RGS dispersion axis measured from north through east. The 90 per cent PSF extraction region corresponds to approximately 70 arcsec (50~kpc) width across the cluster centre. We grouped the spectra to have a minimum of 20 counts per spectral bin and minimised the $\chi^2$ during the spectral fitting. We used the \textsc{rgsbkgmodel} tool to generate background spectra for the spectral fitting. The first and second order data were fit between 7---25 and 7---17~{\AA}, respectively. We fit the first and second order spectra from the two RGS detectors simultaneously, but allowed for constant normalisation offsets between the datasets.

We first used a cooling flow model which assumes that the gas is cooling in steady state through different temperature ranges in the absence of heating (as was applied to the Centaurus cluster; \citealt{SandersRGS08}). We measure the predicted mass deposition rate as a function of temperature, splitting the cooling into temperature ranges of 3.2---2.4, 2.4---1.6, 1.6---0.8, 0.8---0.4 and 0.4---0.08~keV. We use the \textsc{apec} spectral model to generate the cooling spectra. The same Galactic absorption was applied to each component (fitting for the value). The component spectra were smoothed to account for the effect of spectral broadening because of the spatial extent of the source (as in \citealt{SandersRGS08}). We applied different smoothing scales to the upper two temperature range components. The remaining cooler components were smoothed with a single broadening component. We tied the metallicities between each component, fitting for N, O, Ne, Mg, Si, Ca, Fe and Ni. The reduced $\chi^2$ of the final fit was $2658/2468 = 1.07$.

\begin{figure}
  \centering
  \includegraphics[width=0.9\columnwidth]{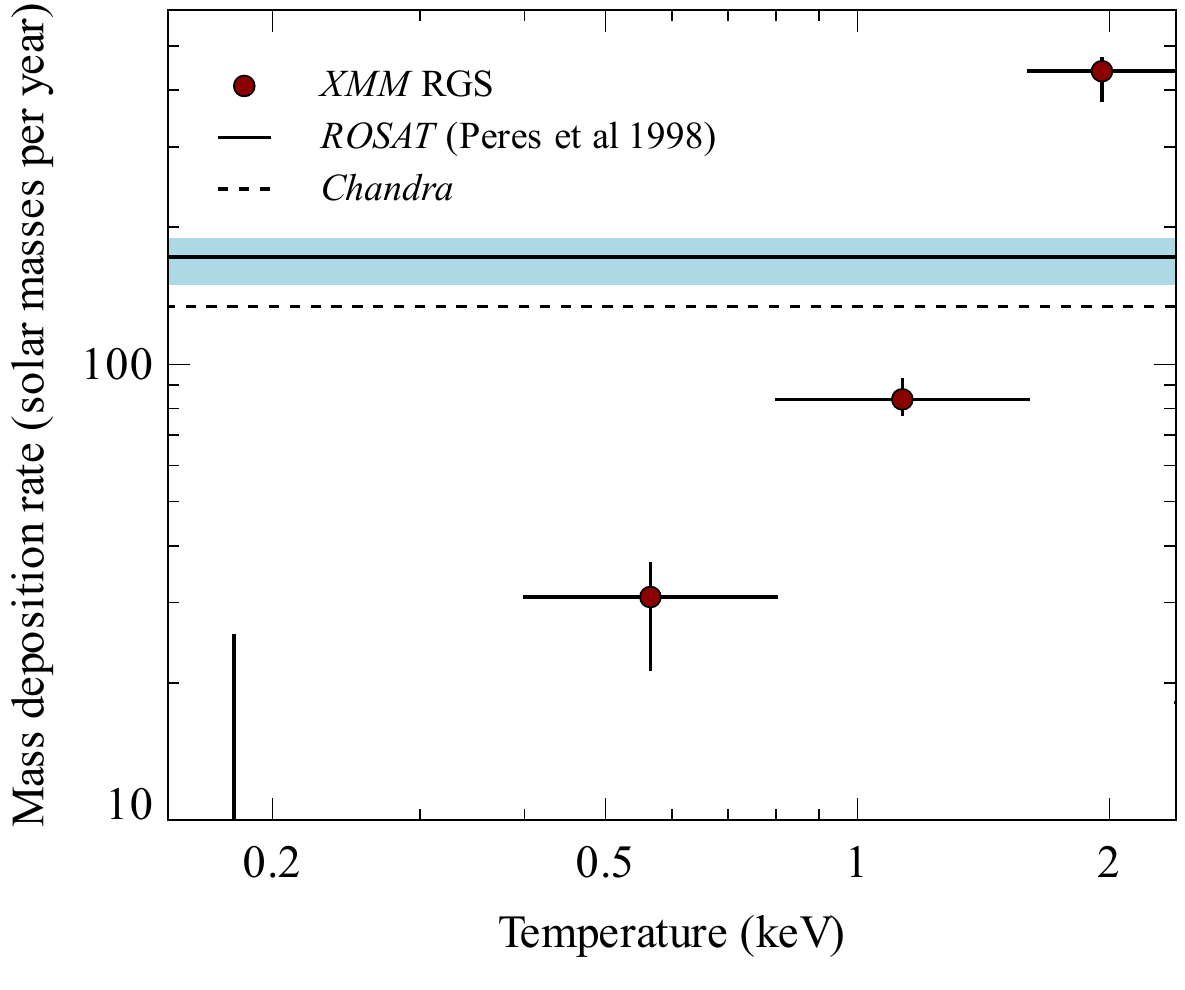}
  \includegraphics[width=0.9\columnwidth]{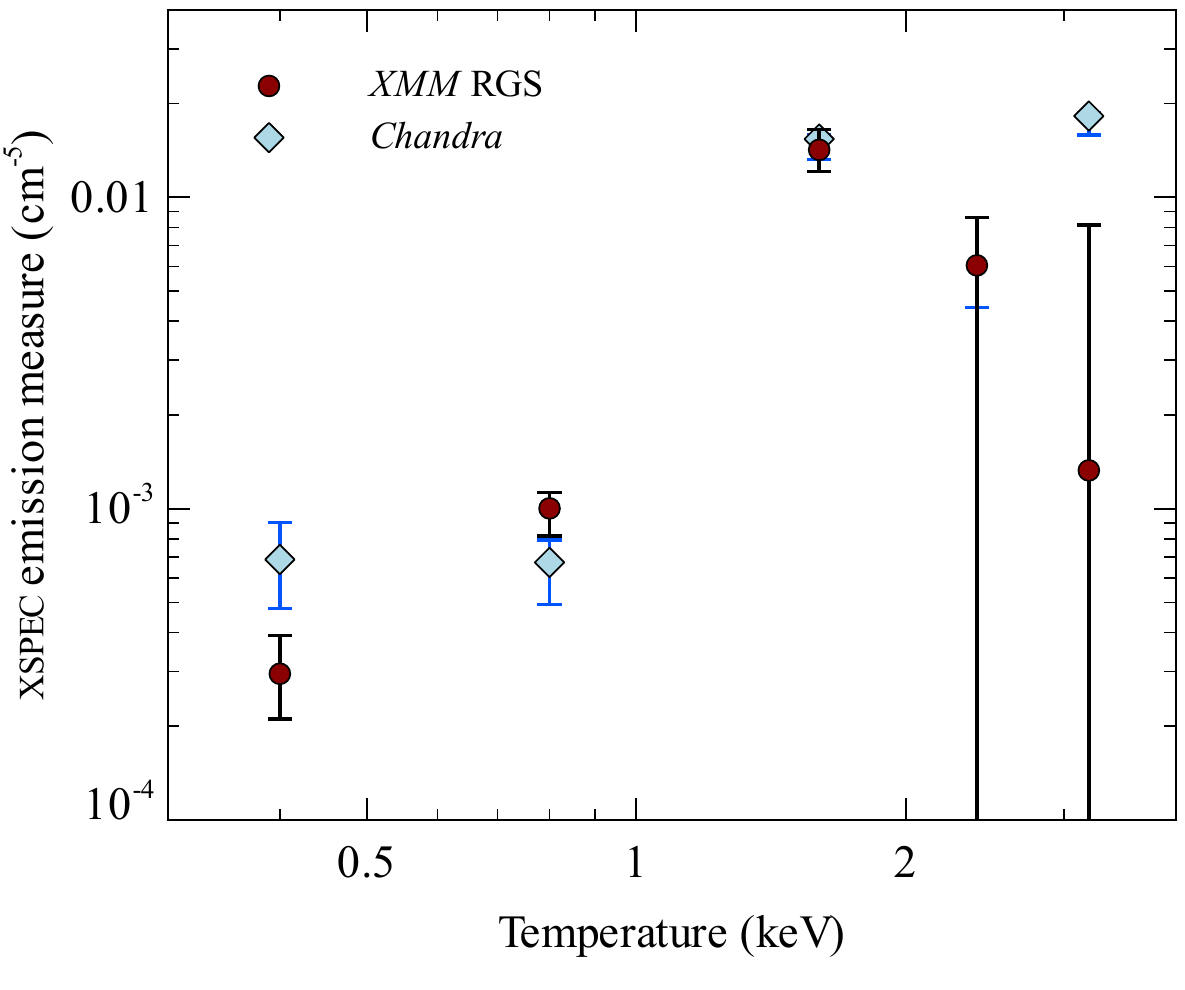}
  \caption{(Top panel) Mass deposition rates obtained from \emph{XMM-Newton} RGS data compared to values from \emph{Chandra} spectral fitting and from \emph{ROSAT} deprojection of luminosity profiles \citep{Peres98}. The shaded regions show the statistical uncertainties in the two mass deposition rates. Below 0.4 keV we find a $1\sigma$ upper limit of $25\Msunpyr$ from the RGS data. (Bottom panel) Best fitting emission measures comparing values between \emph{XMM-Newton} RGS and \emph{Chandra}.}
  \label{fig:rgsmultitemp}
\end{figure}

Fig.~\ref{fig:rgsmultitemp} (top panel) shows the obtained values of the mass deposition rate as a function of temperature. For comparison, we show the mass deposition rate obtained from the surface brightness deprojection of \emph{ROSAT} data \cite{Peres98}, correcting the result to our cosmology. We also show the mass deposition rate from fitting a cooling flow component plus thermal component model to the \emph{Chandra} spectrum extracted from a 70 by 294 arcsec rectangular region rotated to the position angle of the RGS observation,  containing the bright core. The spectrum was fitted between 0.6 and 7~keV. The cooling flow component minimum temperature was fixed at the minimum possible value.

We can compare the high spectral resolution temperature distribution from that obtained with our low spectral resolution \emph{Chandra} data. We fitted a model consisting of five different single \textsc{vapec} thermal components at 3.2, 2.4, 1.6, 0.8 and 0.4~keV, linking the metallicities to be at the same fitted value. We show the obtained emission measures from the \emph{Chandra} region above and the \emph{XMM-Newton} RGS in Fig.~\ref{fig:rgsmultitemp} (bottom panel). The abundances were linked between the different temperature components in the fits. We find reasonable agreement at low temperatures.  Higher temperatures come from a much larger region and we have not properly accounted for the different effective areas of the detectors as a function of position. The RGS is also poorer at detecting hot gas because of its limited wavelength range.

\section{Multiwavelength comparison}
\subsection{H$\alpha$+[N~\textsc{ii}] emission}
\begin{figure}
  \centering
  \includegraphics[width=\columnwidth]{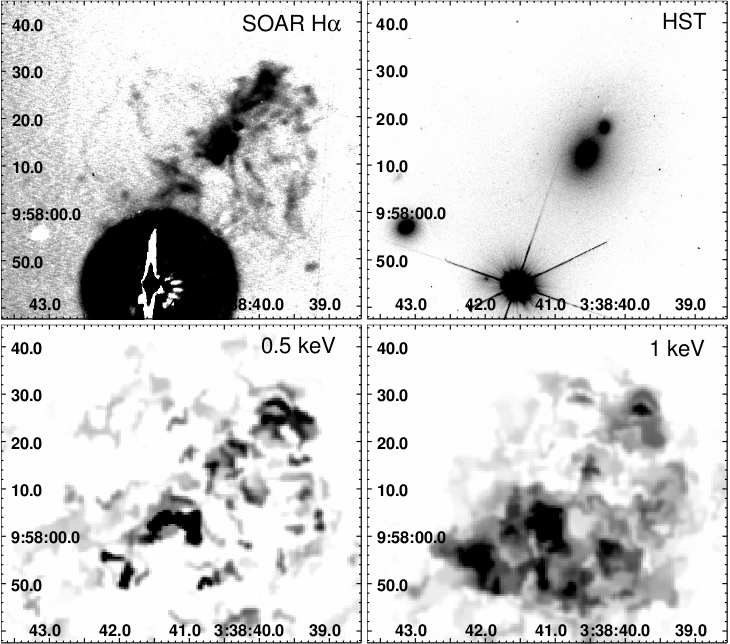}
  \caption{(Top left panel) Narrow band continuum-subtracted H$\alpha$+[N~\textsc{ii}] image from the SOAR telescope \citep{Donahue07}. The black disc to the bottom is a nearby star. (Top right panel) \emph{HST} F606W image of the cluster (association U5A40701B). (Bottom panels) Emission measure per unit area maps from X-ray spectral fitting for 0.5 and 1~keV components, smoothed with a Gaussian of 1 arcsec.}
  \label{fig:hacoolgas}
\end{figure}

The H$\alpha$ nebulae in the cores of galaxy clusters are often correlated with the cool X-ray emitting gas (e.g. Perseus, \citealt{FabianPerFilament03}, Centaurus, \citealt{Crawford05} and A1795, \citealt{Crawford179505}). In Fig.~\ref{fig:hacoolgas} we show a SOAR H$\alpha$ image of the nebula \citep{Donahue07}, an \emph{HST} image of the same region, and emission measure per unit area maps of the 0.5 and 1~keV components. The morphology of the emission measure of the cooler 0.5~keV component matches the H$\alpha$ emission remarkably well. The 1~keV emission measure, which more closely matches the X-ray image, has some correspondence, but contains regions without associated H$\alpha$ flux. We will examine this further in Section \ref{sect:coolblobs}.

\subsection{Radio emission}
\begin{figure*}
  \centering
  \includegraphics[width=\textwidth]{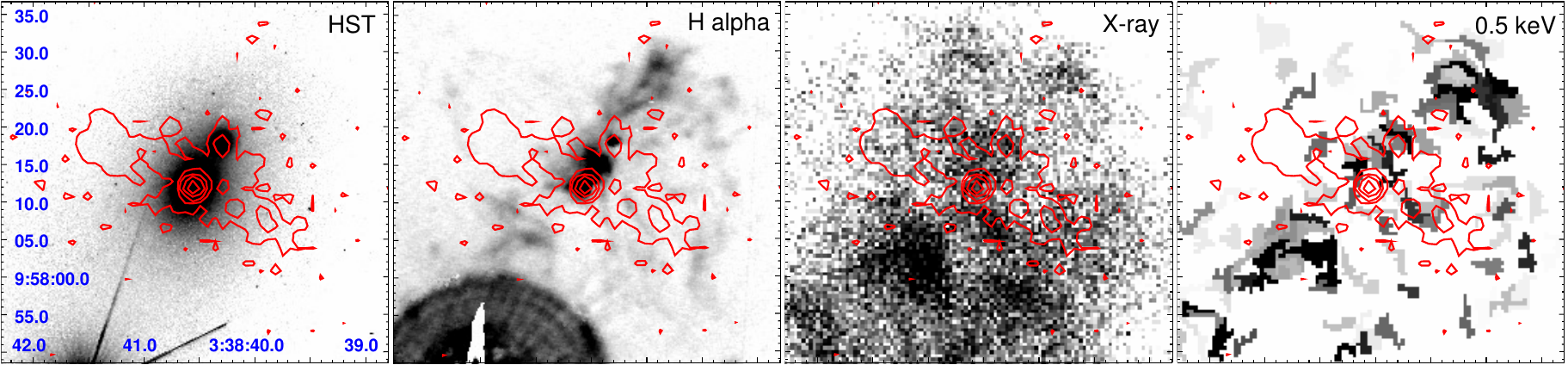}
  \caption{Contours of 5~GHz radio emission shown on images in other wavebands. (Left panel) \emph{HST} F606W image, (centre-left panel) SOAR H$\alpha$+[N~\textsc{ii}] image, (centre-right panel) 0.5 to 5~keV \emph{Chandra} image and (right panel) 0.5~keV emission measure per unit area map.}
  \label{fig:6cmcontours}
\end{figure*}

\begin{figure}
  \includegraphics[width=\columnwidth]{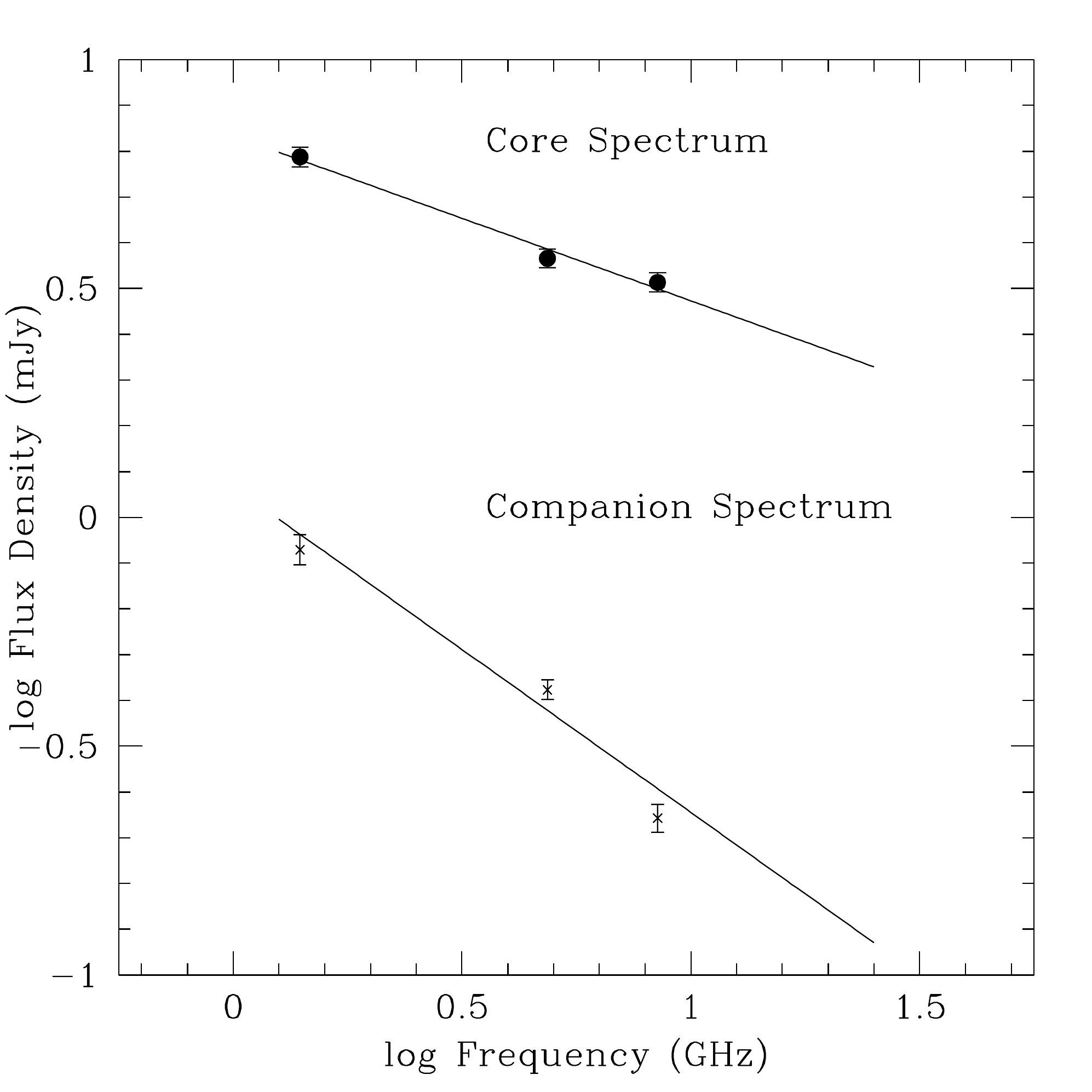}
  \caption{Radio spectrum of the nucleus of the main galaxy and its companion to the north west. The core of the main galaxy has a spectral index of $-0.36 \pm 0.04$, and the companion has a spectral index of $-0.71 \pm 0.06$.}
  \label{fig:radiospec}
\end{figure}

We obtained new VLA observations of the 2A~0335+096 cluster, and combined these data with observations from the NRAO archive (some of which has been previously published by \citealt{Sarazin2A033595}).  The new VLA observations were obtained on 2006 November 21 at 5~GHz when the VLA was in its C configuration.  These observations were combined with B and D configuration observations to form a moderate resolution ($1.49 \times 1.43$~arcsec) image with good sensitivity to extended emission (rms noise 21 $\mu$Jy).  In the 5~GHz image (Fig.~\ref{fig:6cmcontours}) we see a compact core with a peak of 3.3 mJy.  Faint jets extend to the northeast and southwest for $\sim 12$~arcsec in each direction. In addition there is a compact radio source detected, with flux density $0.54\pm0.02$~mJy,  coincident with the optical nucleus of the companion galaxy (Fig.~\ref{fig:6cmcontours}).  We show in Fig.~\ref{fig:radiospec} the radio spectrum of the nucleus of the central galaxy and its companion to the northwest.

We imaged the compact core of 2A~0335+096 using the VLBA at 5~GHz on 2007 Jan 5. Because of the low flux density of the compact core we performed the observation in a phase-referencing mode, alternating short observations of 2A~0335+096 with the nearby calibrator J0334+0800. The compact core is detected at a position of (J2000) RA 03:38:40.55047, Dec $9^\circ58'12.0415''$ with an uncertainty of 0.8~mas in each coordinate.  The core remains compact on the mas scale, having a size of just 2~mas.  No evidence for a jet is seen and the flux density of the core measured by the VLBA is $3.5 \pm 0.8$~mJy, indicating that the VLBA recovers all the flux detected by the VLA, and there is no significant jet emission on scales from 3~mas to 0.3~arcsec.  On larger scales the VLA does detect the jet as reported above.  The location of the nucleus of the companion galaxy was searched, but not detected by our VLBA observations with a $5 \sigma$ noise limit of $<0.5$~mJy.  Given the low flux density of the radio source this result is not conclusive.

\begin{figure}
  \centering
  \includegraphics[width=0.8\columnwidth]{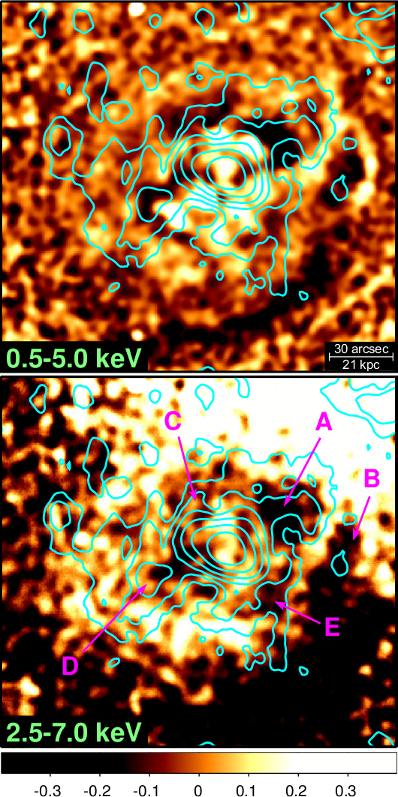}
  \caption{Comparison of 1.5~GHz radio emission (in contours), (top panel) 0.5 to 5~keV X-ray unsharp masked image, showing the ratio difference between an image smoothed by a Gaussian of 0.5 arcsec and 14.8~arcsec, then smoothed and (bottom panel) 2.5 to 7~keV image, showing the fractional difference from the average at each radius, then smoothed. The colour bar shows the fractional differences. The contours are spaced logarithmically, with 9 contours between $10^{-4}$ and 0.011 Jy~beam$^{-1}$. The depressions in X-ray surface brightness are labelled as in Fig.~\ref{fig:unsharp}, with an additional depression E.}
  \label{fig:20cm}
\end{figure}

At 1.5~GHz the radio emission is much more extended than at 5~GHz (Fig.~\ref{fig:20cm}), showing a mini halo. Only a few mini halos have been detected, and all of them appear to be associated with cooling core clusters such as 3C84, A1835, A2029 and Ophiuchus \citep{Govoni09}. Although the radio emission strongly peaks towards the centre of the cluster, there are radio enhancements associated with some or all of the depressions in the X-ray surface brightness, as seen in the unsharp-masked 0.5 to 5~keV image in the left panel of Fig.~\ref{fig:20cm}, in particular cavities A and D. The X-ray cavities are easier to see in a harder band image of the cluster (rightmost panel), which is more sensitive to the X-ray thermal pressure \citep{FormanM8707} and eliminates much of the emission from cool X-ray emitting gas.

In addition we see a depression E which was not so obvious in Fig.~\ref{fig:unsharp} because of the nearby cool X-ray emitting gas. This depression is associated with a spur in the 1.5~GHz emission.

\begin{figure}
  \centering
  \includegraphics[width=0.9\columnwidth]{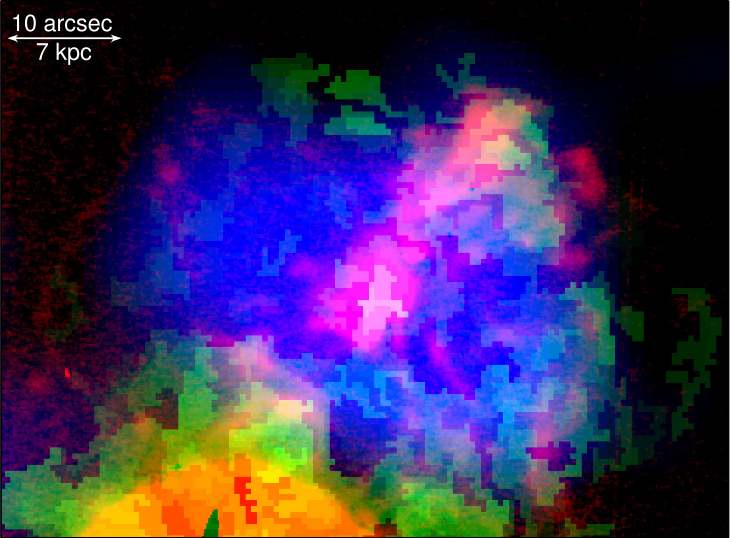}
  \\
  \includegraphics[width=0.9\columnwidth]{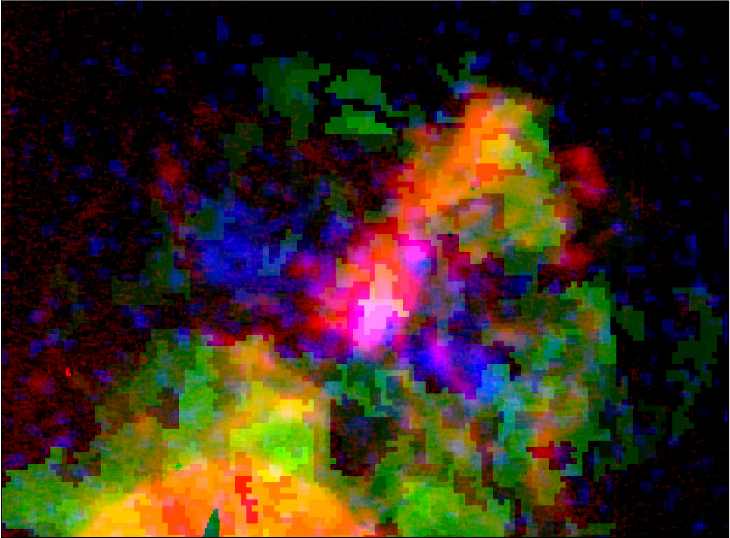}
  \caption{RGB images showing the correlation between H$\alpha$+[N~\textsc{ii}] (red), 1 keV temperature emission measure (green) and radio (blue). The top panel shows the 1.5~GHz radio emission, while the bottom panel shows 5~GHz. The top of the bright red circle to the bottom left is a foreground star.}
  \label{fig:rgb_ha_xray_radio}
\end{figure}

In Fig.~\ref{fig:rgb_ha_xray_radio} we show overlays of the H$\alpha$ (red), 1 keV emission measure (green) and radio (blue, the top panel showing 1.5~GHz, while the bottom panel shows 5~GHz). The strongest 1.5~GHz emission and H$\alpha$ emission appears to be bounded by the region of 1~keV temperature gas. The 5~GHz jets also appear to lie within this same region.

\section{Discussion}
\subsection{Cool X-ray emitting blobs}
\label{sect:coolblobs}
We are able to shed new light on the nature of the cool X-ray emitting blobs in the core of the cluster. The regions require multiple temperature components to fit their X-ray spectra (Fig.~\ref{fig:multitemp}), containing temperatures between $\sim 0.5$ and 2~keV. If the different components are in pressure equilibrium then this requires that the volume filling fraction of the lower temperature components is much lower than those at higher temperatures. The cool X-ray emitting gas is likely to be similar in nature to that found associated with the emission-line filaments in Perseus, which are calculated to have a small volume filling fraction \citep{SandersPer07}.

\cite{Mazzotta03} and \cite{Donahue07} found that the dominant X-ray emission was not similar in morphology to the optical line emission. However we find that the lower temperature 0.5~keV gas has a morphology much closer to the H$\alpha$ map. The multi-temperature spectral model is able to separate the different gas phases and show there is association at low X-ray temperatures.

The presence of cool X-ray emitting gas in this cluster with a temperature around 0.5~keV is confirmed by Fe~\textsc{xvii} emission lines in high resolution grating spectra (Fig.~\ref{fig:rgsspec}; \citealt{Werner06}). We quantify the amount of emission from cold material with \emph{XMM-Newton} RGS and \emph{Chandra}, showing reasonable agreement between the two measurements. The thermal electron pressure in the core is around $0.07 \keVpcmcu$. If the cold X-ray emitting gas is in pressure equilibrium, the emission measure at 0.4~keV implies it has a volume of $3 \times 10^{66} \cmcu$. It could be contained within a cube of size $4.7 \kpc$. Therefore the volume filling factor is low as the spatial extent of the emission is much larger (Fig.~\ref{fig:multitemp}), unless it is sheet-like. The volume filling factor of this cool material in the core (within a radius of 22 kpc) is 0.2~per~cent.

The range in X-ray temperatures in this cluster is at least a factor of 8. Similarly large ranges of temperature are seen in other nearby cool core clusters with good X-ray data (e.g. Centaurus, \citealt{SandersRGS08}; Perseus, \citealt{FabianPer06}; and Abell 2204, \citealt{SandersA220409}). At lower temperatures, successively less material is seen than would be expected from simple cooling (as seen in the sample of \citealt{Peterson03}).

The RGB images in Fig.~\ref{fig:rgb_ha_xray_radio} show that the 1~keV component, the H$\alpha$ flux and radio emission are connected. There is a `bubble' around 10~kpc in radius around the nucleus of the central galaxy. In this region there is no 1~keV X-ray emitting material (it surrounds this region), but there is strong radio and line emission. The X-ray gas is cooler in this bubble and there is cool line emitting gas, but there are also non-thermal particles associated with the jets and the minihalo. The cool 0.5 keV component and H$\alpha$ lies roughly perpendicular on the sky to the 5~GHz radio emission.

\subsection{The low temperature spiral}
Our temperature maps show an obvious spiral of low-temperature X-ray emitting gas (at least in terms of projected emission-weighted temperature), on scales of an arcminute to several arcminutes. This adds to the list of galaxy clusters showing such spirals, including Perseus \citep{FabianPer00,ChurazovPer03,SandersPer04,FabianPer06}, Centaurus \citep{SandersCent02,Fabian05}, Abell 2029 \citep{Clarke04} and Abell 2204 \citep{SandersA220405,SandersA220409}. A spiral morphology is also seen in surface brightness in the Ophiuchus cluster \citep{Ascasibar06}.

Spiral like features are often associated with the presence of cold fronts. \cite{Ascasibar06} generate spirals and cold fronts in smooth particle hydrodynamical (SPH) simulations of galaxy clusters. The features are caused by the interaction of small subhalos with a primary cluster. The temperature maps created by their simulations are similar to what is observed in this cluster. Grid based codes do not tend to see the long-lived large scale structures seen by SPH codes \citep{Ascasibar06}, which is probably due to the lack of artificial viscosity \citep{Dolag05}. As pointed out by \cite{Ascasibar06}, the similarity of real clusters with SPH cold fronts and spirals may indicate that viscosity is significant in galaxy clusters. The straightness of the emission line filaments in Perseus \citep{FabianPerFilament03} also indicates that viscosity may be important in the intracluster medium.

The spiral shape seen in metallicity matches the temperature spiral in some respects (Fig.~\ref{fig:zmap}), although there is not an exact one to one correspondence between the lower temperature gas and higher metallicity material. The correspondence supports the idea that the spiral is caused by the interaction of two subhalos.

\subsection{The surface brightness edge to the south west}
\cite{Mazzotta03} previously reported a surface brightness discontinuity to the south of the cluster core. They observed no temperature jump associated with the edge, but found a density jump of a factor of 1.6. They interpreted this edge as a cold front (see \citealt{MarkevitchCFShock07} for a review). A static cold front should show no pressure discontinuity across it. The pressure change by a factor of $1.6 \pm 0.3$ was therefore interpreted as evidence that the cool core was moving with respect to its surroundings with a Mach number of $0.75 \pm 0.2$.

\cite{SandersShock06} instead suggested that the lack of a temperature jump indicated that the edge was an isothermal shock as found in the Perseus cluster. In Perseus are seen high pressure rings surrounding the inner X-ray cavities, presumably weak shocks caused by the inflation of the cavities. However, they have no associated temperature increase with respect to their surroundings. In Perseus it appears that the only currently understood mechanism which could suppress the temperature change is mixing of cool gas associated with the emission line filaments with the shocked gas \citep{Graham08Per}.

Our new results are not consistent with either a fast-moving subcore or an isothermal shock (though note we use a different range of angle to \citealt{Mazzotta03}). Fig.~\ref{fig:coldfront} shows no pressure discontinuity associated with the edge.  Fitting a line through the two points inside and two points outside the edge, we calculate that there can be a 26 per cent maximum pressure change associated with it (at the $2\sigma$ confidence level). This translates into a Mach number upper limit of 0.55 (examining figure 6 in \citealt{Vikhlinin01}) of the cool subcore with respect to its surroundings. Taking a gas temperature of 3.5~keV, this translates into a maximum velocity of $540 \kmps$.

\subsection{X-ray cavities}
We detect five clear X-ray cavities surrounding the cluster core, two of which were previously detected \citep{Sarazin2A033595,Mazzotta03}. Cavities A and D (Fig.~\ref{fig:unsharp}) show 1.5~GHz radio emission associated with them (Fig.~\ref{fig:20cm}), although the radio contours blend into the surrounding minihalo. Cavity C is much less clear, but the radio emission has some kinks near its position. Cavity B, obvious from the surrounding bright X-ray emission, has some low significance radio contours within it. Cavity E was not so obvious at soft energies, but is seen in a high energy X-ray band and shows a spur of radio emission pointing along it. Most of the extended radio emission appears to be associated with X-ray cavities. This indicates that the minihalo is mostly formed from episodes of AGN activity.

There is not a clear anticorrelation between the 5~GHz radio jet and the X-ray emission. This may be because the jet has entrained X-ray emitting gas, or the jet is emitting in the X-ray waveband itself.

\begin{table}
  \caption{Estimated radii for each of the cavities. Also shown are the computed volumes and $PV$ energies, where $P$ is the total thermal pressure and $V$ is the bubble volume.}
  \begin{tabular}{llll}
    Cavity & Radius (arcsec) & Volume ($10^{3} \kpc^{3}$) & $PV$ ($10^{58}$ erg)\\ \hline
    A      & 18              & 8.3                        & 4.1 \\
    B      & 9               & 1.0                        & 0.45\\
    C      & 14              & 3.9                        & 1.9 \\
    D	   & 17              & 7.1                        & 3.5 \\
    E      & 12              & 2.5                        & 1.2 \\
  \end{tabular}
  \label{tab:cavities}
\end{table}

We can estimate the volume of the cavities if we assume they are spheres (see Table \ref{tab:cavities}). Cavity E may have a morphology closer to a spherical-cap bubble, however. Cavities A, C, D and E lie at roughly the same radius from the centre of the cluster as each other. We can estimate the electron pressure at the radii of the cavities by deprojecting spectra from annuli using the direct spectral deprojection technique. We obtain electron pressures of around $0.057 \keVpcmcu$ at this radius. The electron pressure at the radius of B is similar ($0.052 \keVpcmcu$). The total thermal pressure at the bubble times the volume of the bubble gives an estimate of its energy content. The work that can be done by a rising bubble on its surroundings is its enthalpy, for a relativistic gas $4PV$ (e.g. \citealt{Birzan04}; \citealt{DunnFabian06}). In Table~\ref{tab:cavities} we show the quantity $PV$ which can be used to estimate the amount of heat which can be injected by each bubble.

The total $4PV$ enthalpy of these cavities would be on the order of $5 \times 10^{59} \erg$. The luminosity from within the 122~kpc cooling radius of this cluster is $2.9 \times 10^{44} \ergps$ \citep{Birzan04}. The heating power of these bubbles would be sufficient to heat the cluster core for $\sim 5 \times 10^{7} \yr$.

The morphology of the cavities is complex. It is unclear what the order of generation of the bubbles was. The fact that cavity B is much weaker in radio may indicate it is a ghost cavity and is significantly older than the others. Cavity B also lies at the edge of a bulge of cool emission that appears extended in its direction. The bulge behind may be gas pulled up in the wake of the bubble. The cavity also has a region of bright emission `above' it. Fig.~\ref{fig:colimages} (bottom panel) also indicates that cavities A and B may be halves of a single cavity. In such a complex system, projection effects are very important in interpreting what we see and what the total energy budget is.

\section{Summary}
Our deep \emph{Chandra} X-ray observation of 2A~0335+096 shows a wealth of structure in the core, including cool bright blobs, a cool metal rich spiral, multiple cavities and a cold front. The spiral may indicate that the core of the system has undergone an interaction with a smaller subhalo.

Most or all of the cavities are associated with 1.5~GHz radio emission, showing the nucleus has been active in sending radio bubbles into its surroundings. The $4PV$ enthalpy in the cavities would be sufficient to heat the cluster core for around $5 \times 10^7 \yr$.

The blobs of bright X-ray emission previously seen in the core of this cluster contain multiphase X-ray emitting material. The coolest observable material, at around 0.5~keV temperature, is correlated well with the striking H$\alpha$+[N~\textsc{ii}] emitting nebula. There also appears to be a complex interaction between the H$\alpha$ nebula, 1~keV X-ray gas and radio emitting plasma.

We find no evidence for a isothermal shock across the surface brightness edge or for a moving subcore in the cluster. From the lack of pressure discontinuity, we find an upper limit of a Mach number of 0.55 ($2\sigma$) for the relative velocity of cool core and surroundings.

\section*{Acknowledgements}
ACF thanks the Royal Society for support. GBT acknowledges support for this work from the National Aeronautics and Space Administration through Chandra Award Number GO7-8124X issued by the Chandra X-ray Observatory Center, which is operated by the Smithsonian Astrophysical Observatory on behalf of the National Aeronautics and Space Administration under contract NAS8-03060A.

The National Radio Astronomy Observatory is a facility of the National Science Foundation operated under a cooperative agreement by Associated Universities, Inc.

We thank M.~Donahue for kindly providing the SOAR image of the emission line filaments.

\bibliographystyle{mnras}
\bibliography{refs}

\clearpage
\end{document}